\documentclass[12pt,preprint]{aastex}

\def\kms{km~s$^{-1}$}
\def\degpnt{^{\circ}\kern-1.7mm.\kern+.35mm}
\def\arcpnt{"\kern-1.7mm.\kern+.35mm}
\def\minpnt{'\kern-1.0mm.\kern+.30mm}

\def\lsun{L$_\odot$}

\begin{document}


\title{Tramp Novae Between Galaxies in the Fornax Cluster: Tracers of
	Intracluster Light}


\author{James D. Neill}
\affil{Department of Physics and Astronomy, University of Victoria, 
	Elliott Building, 3800 Finnerty Road, Victoria, BC, V8P 1A1, Canada}
\email{neill@uvic.ca}

\author{Michael M. Shara}
\affil{American Museum of Natural History, 79th and Central Park West
       New York, NY, 10024}
\email{mshara@amnh.org}

\author{William R. Oegerle}
\affil{Laboratory for Astronomy and Solar Physics, NASA/GSFC, 
	Greenbelt, MD 20771}
\email{oegerle@uvo.gsfc.nasa.gov}



\begin{abstract}

We report the results of a survey for novae in and between the galaxies of
the Fornax cluster.  Our survey provides strong evidence that intracluster
novae exist and that they provide a useful, independent measure of the
intracluster light in Fornax.  We discovered six strong nova candidates
in six distinct epochs spanning eleven years from 1993 to 2004.  The data
were taken with the 4m and the 1.5m telescopes at CTIO.  The spatial
distribution of the nova candidates is consistent with $\sim$16-41\%
of the total light in the cluster being in the intracluster light, based
on the ratio of the number of novae we discovered in intracluster space
over the total number of novae discovered plus a simple completeness
correction factor.  This estimate is consistent with independent measures
of intracluster light in Fornax and Virgo using intracluster planetary
nebulae.  The accuracy of the intracluster light measurement improves
with each survey epoch as more novae are discovered.

\end{abstract}




\section{Introduction\label{fnx_intro}}

A fundamental problem in understanding the evolution of galaxy clusters
and their constituent galaxies is the importance of dynamical processes
such as tidal stripping.  Theoretical models of cluster evolution have
considered collisions, mergers, dynamical friction and tidal stripping in
order to explain the galaxy morphology-density relation and the presence
of luminous halos around supergiant D galaxies.  It is generally accepted
that tidal stripping will remove the more loosely bound stars in the
outer envelopes of galaxies \citep{ric76}.  \citet{mer84} has argued
that the extensive envelopes of cD galaxies consist of this tidal debris
moving in the potential well of the cluster.  However, the efficiency
of tidal stripping is not well understood, because of uncertainties in
the velocities, the mean tidal field, the distribution of dark matter,
and the orbits of stars in the outer envelopes of the stripped galaxies.
Typical theoretical estimates indicate that 10 - 20\% of the total
cluster light should be in the intergalactic medium \citep{wil04,mur04} 
and that the
amount of ICL should increase with time \citep{wil04}.  \citet{mil83}
and \citet{dre84} have therefore stressed the importance of obtaining
reliable measurements of the intracluster light (ICL) as a direct
indicator of the tidal damage suffered by galaxies.

Attempts to measure the ICL date back to the early 1970s, originally
being done with photographic plates.  More recently CCD surveys have
been employed by \citet{par90}, \citet{bou90}, \citet{vil94}, and 
\citet{ber95}
with results ranging from non-detections up to $\sim50$\% of the total
cluster light (surface brightnesses of $\mu_V \sim29$ mag arcsec$^{-2}$).
Even with CCDs, detecting the ICL directly is notoriously difficult and
fraught with uncertainties such as the contributions from overlapping
galaxy halos.  An independent means of measuring the ICL, not subject
to the same problems as measuring the extremely low surface brightness
diffuse emission is needed.

Methods using planetary nebulae (PNe) have been developed and applied
to Virgo \citep{fel98,arn03,fel03} and Fornax \citep{the97}.  Intergalactic
red-giant-branch stars have also been detected in Virgo by comparing the
luminosity function of point sources in intracluster space with control
fields far from any cluster or galaxy \citep{fer98,dur02}.  Intracluster
SNe have also been identified and have extended measurements of the ICL 
out to $z = 0.1$ \citep{gal03}.

Novae typically range in brightness from M$_V = -6$ to $-10$, while the
bright end of the PN luminosity function reaches M$^*_{5007} = -4.5$
\citep{cia89} and red-giant-branch stars typically reach M$_V = -3$.
Supernovae are much brighter, but also much rarer than novae.
Novae can therefore be detected more easily in a given cluster, or
be observed to greater distances than PNe or red-giant stars.
The transient nature of novae and their H$\alpha$ brightness help eliminate
the contamination suffered by PNe or red-giant-branch stars due
to background emission line objects or unresolved compact galaxies.
Another appealing property of novae is that the rate of novae per unit mass
is high enough such that with a deep enough survey, one can expect to
accrue a statistically interesting number of novae in each epoch, thus
increasing the accuracy of the ICL measurement with each new epoch.

These properties give novae the potential to be useful as an independent
and accurate measure of the ICL in nearby clusters.  In spite of this, to
our knowledge, no intra-cluster novae have been reported in the
literature to date.  As an independent check on highly uncertain direct 
measures of intracluster light, we can use novae to
verify the amount of stellar tidal debris in the cluster.  This then
places constraints on the importance of tidal stripping in the formation
and evolution of galaxy clusters.  Our M81 nova results from the most
complete nova survey ever undertaken
indicate that novae are preferentially associated
with the older, bulge populations in galaxies \citep{nei04}.  We can
use this knowledge to attempt to discern the predominant source of the
ICL either in bulge-dominated or disk-dominated galaxies.

Another lesson we have learned from our nova studies is that accurate
rates and distributions can only be derived from comprehensive,
densely time-sampled surveys \citep{nei04}.  The data set we use in this
paper is diverse and very sparsely sampled.  We therefore restrict our
investigation to a relative comparison of galaxy versus inter-galaxy
novae.  Our aim is to prove the {\it feasibility} of using novae to perform
an accurate assessment of the amount and origin of the ICL.

We have acquired six epochs of survey data of the central 30' square
of the Fornax cluster.  Fornax is nearby and has a fairly high surface
density of galaxies in the core, increasing the likelihood of tidal
stripping.  At the distance of Fornax, $(m-M)_0$ = 31.35 \citep{kis99},
novae will have an average m$_B$(peak) $\sim23.35$ (the brightest novae
will have peak brightnesses of m$_B$ = 21.35), and are easily detectable.
Novae have been observed by \citet{pri87} in Virgo cluster galaxies,
which are at the same distance as Fornax.

Novae are very bright in H$\alpha$ (B $-$ H$\alpha \sim0.5$ when the
nova erupts, and B $-$ H$\alpha \sim3$ shortly thereafter).  Furthermore,
novae remain bright in H$\alpha$ for an extended period of time (months),
providing a useful means of discovering them.  To a limiting magnitude of
M$_{H\alpha}$ = -7.5 (m$_{H\alpha}$ = 23.85 at Fornax) the mean H$\alpha$
visibility lifetime is 3 months \citep{cia90a}.  Assuming a nova rate
of 24 yr$^{-1}$ per 10$^{10}$ \lsun(B) \citep{cia90b}, a luminosity
in galaxies of M$_B \sim-22.5$ in the core of Fornax \citep{fer89},
a visibility lifetime in H$\alpha$ of 3 months, and a conservative
completeness factor of 0.5 due to the high surface brightness of galaxy
cores, we estimate that at any given epoch there should be $\sim50$
novae in eruption (with visible H$\alpha$ emission) in galaxies in the
central 1 deg$^2$ of the Fornax cluster.  Since we are surveying the
central $1/2$ deg$^2$ we multiply this estimate by $1/4$ and predict we
will see 12 novae in any given epoch.

\section{Observations}

The observations that went into this study are a diverse set acquired
over the span of more than a decade.  The central 30' of Fornax,
centered on RA 03:38:13.72, Dec -35:32:52.2 (J2000), was surveyed.
This area includes the central dominant (cD) galaxy NGC 1399 and three
other elliptical galaxies: NGC 1404, NGC 1387, and NGC 1389.  This region
was imaged twice in 1993, once in 1995, for five nights in 1999, and once
each in 2003 and 2004.  The data were acquired during programs for other
projects and during engineering time or director's discretionary time.
These observations are summarized in Table~\ref{tabforobs}.

Because of the long-term nature of the data acquisition, there was no
uniform observing strategy employed.  We started with the Tek2048 camera,
and when the MOSAIC II camera became available (before the 1999 run),
we switched to that camera for the remaining epochs.  We used standard 
Johnson B- and R-band filters and narrow-band H$\alpha$ filters primarily,
but for two epochs (b and c, see below) we were constrained to using
a T1 filter from the Washington system ($\lambda_{eff} = 6330$\AA, 
FWHM $=800$\AA, Canterna 1976), which is roughly equivalent to a Johnson R
filter ($\lambda_c = 6440$\AA, FWHM $=1510$\AA).  The H$\alpha$ filter 
used for the 1993 a epoch was the 6600/75 CTIO filter with $\lambda_c = 
6593$\AA\ and $\Delta\lambda = 64$\AA\ which accounts for the velocity
of Fornax (1517 $\pm$ 91 km s$^{-1}$, Grillmair et al. 1994).  The 
H$\alpha$ filter used for the 1999 and 2003 epochs was 
the MOSAIC II c6009 filter with $\lambda_c = 6563$\AA\ and $\Delta\lambda 
= 80$\AA.  There is no H$\alpha$ filter in the MOSAIC II set that
correctly accounts for the velocity of Fornax, so we chose the filter 
whose central wavelength comes the closest to 6600\AA.  We will discuss the
impact of using this filter below.

Total exposure times were restricted by the gaps available in the schedule 
for other programs.  Individual exposure times were chosen to provide at
least three frames per epoch so cosmic rays could be removed in the combining
process.  For the H$\alpha$ frames the only exceptions to this are the 1999 g
epoch (1 exposure) and the 1999 h epoch (2 exposures).  Since cosmic rays
could not be adequately removed from these exposures, they were only used for
nova confirmation and measurement and not detection (see below).

All of the data were acquired using the CTIO 4m telescope with the 
exception of the b and c epoch data (taken exclusively through the T1 
filter) which were acquired with the CTIO 1.5m telescope.

The data acquired in 1993 and 1995 were taken with the Tek2048 chip.
The focal ratios of the 4m and the 1.5m are such that they produce
an identical pixel scale for these images of 0\arcpnt43 per pixel.
The resulting field of view is 14\minpnt7 and thus required four pointings
to cover the central 30' of the cluster.  The seeing tabulated in 
Table~\ref{tabforobs} for these epochs represents the range of seeing 
across the four pointings.  Sky flats were acquired in each filter 
to remove the instrumental response.  

The data acquired in 1999, 2003, and 2004 were taken with the MOSAIC II
camera on the 4m telescope.  This camera is a mosaic of eight 2048x4096
pixel CCDs with a pixel scale of 0\arcpnt27 per pixel.  The resulting
field of view is 36\minpnt8 and thus required only one pointing.
Sky flats were used to remove the instrumental response for the 1999
epochs, and dome flats were used for the 2003 and 2004 epochs.

All data were reduced using the standard tools in IRAF \citep{tod86}
or the {\bf mscred} external IRAF package developed by the NOAO MOSAIC
instrument team.  Individual exposures for a given epoch were combined
to produce a coadded image for the epoch and to remove all but a few
cosmic rays.

\section{Nova Detection \label{novae_det}}

Classical novae erupt with amplitudes of up to 20 magnitudes, reaching
absolute magnitudes as high as M$_B = -10$ and fading back to their
pre-outburst brightness in a matter of weeks or months \citep{war95}.
Novae in the Fornax cluster are therefore
invisible before outburst and fade back to invisibility after outburst
on time scales of less than a year.  Thus, our primary criterion for
detecting novae is that they be transient on this timescale.  For the
data used in this study this means that they must only appear in only
one year's images.  If a variable object appears in images from more
than one year, it is most likely some other kind of variable source:
a foreground periodic, or semi-periodic variable, or a background,
unresolved active galaxy nucleus (AGN).

Classical novae also exhibit strong H$\alpha$ emission in outburst,
providing the next criterion: the candidate should be H$\alpha$-bright.
Data from 1993 and 1999 were taken both through H$\alpha$ and red
continuum filters allowing us to produce H$\alpha$ - red continuum
subtracted images.  These images were used to test candidates from 1993
and 1999 against this criterion.

The primary method for implementing these criteria consisted of performing a
spatial registration of the different epochs, and then blinking them
against each other.  We required that any
candidate nova be visible on each of the individual exposures for the
epoch in which it was detected.  This eliminated cosmic ray hits from
our candidate list.  This method worked well for candidates outside of
the galaxies.  For candidates in the galaxies, we employed the spatial
median filter technique described in \citet{nei04} with a box size of
17 pixels.  The median subtracted image was then blinked in the same
way as the unsubtracted images.

For the epochs taken through an H$\alpha$ - continuum filter pair, we
scaled and subtracted the continuum image from the H$\alpha$ image to
accentuate the objects with H$\alpha$ emission.  The a (1993) and the e,
g, and h (1999) epochs had this filter combination which allowed us to
blink the H$\alpha$-bright objects in these epochs.

\begin{figure}
\plotone{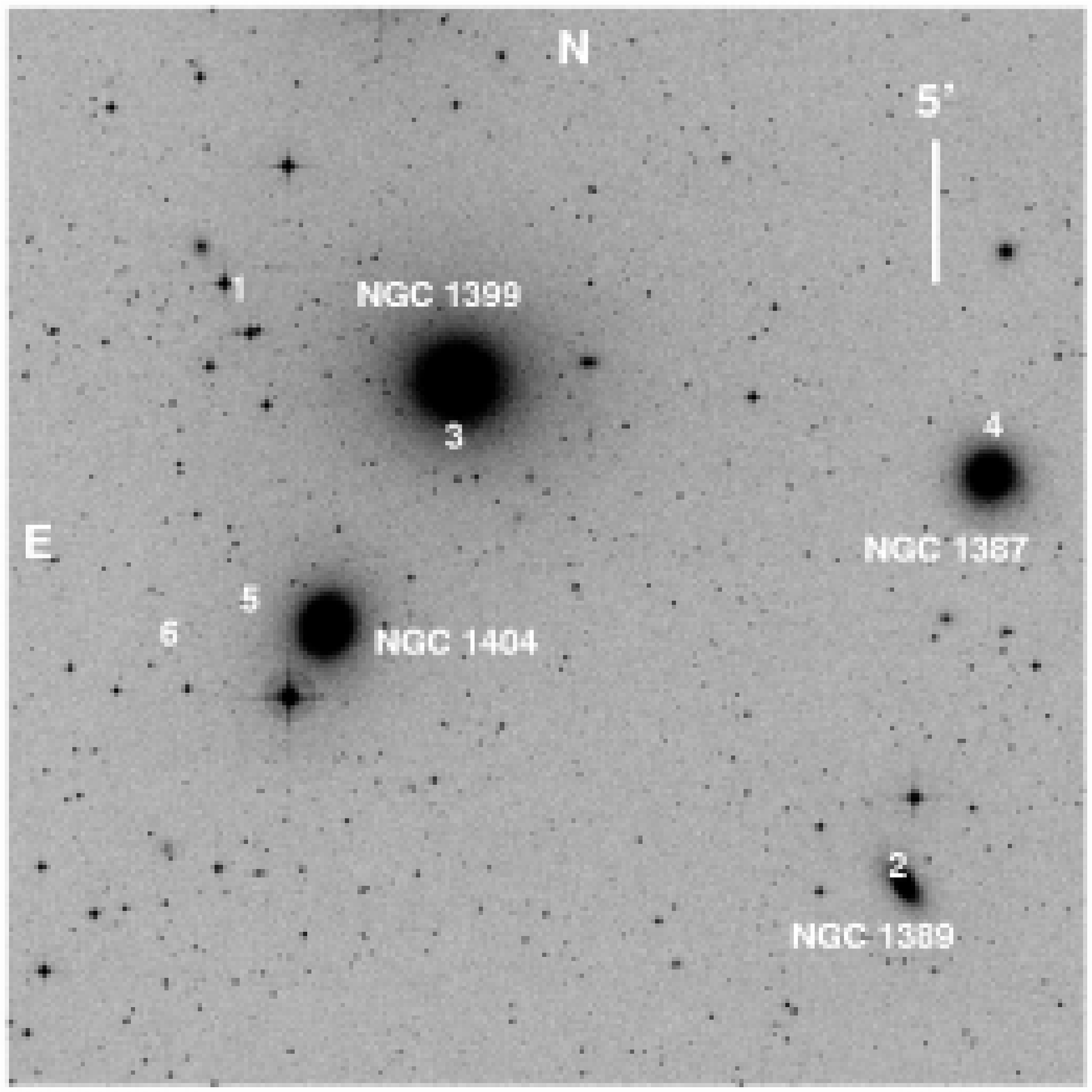}
\caption{Locations of the 6 novae discovered in the Fornax cluster
overplotted on a DSS image. Novae are indicated by their
identification number.  North and
East are at their usual orientation as labeled.  The field of view is 38'
on a side.  The galaxy identifications are indicated, along with a 5'
scale bar in the upper right corner.  Note the positions of Nova 1 and
Nova 6, strong intracluster tramp candidates.
\label{fnx_nov_pos}
}
\end{figure}

Figure~\ref{fnx_nov_pos} shows the locations of the detected nova
candidates on an image from the Digitized Sky Survey\footnote{ The
Digitized Sky Survey was produced at the Space Telescope Science Institute
under U.S. Government grant NAG W-2166.  The images of these surveys are
based on photographic data obtained using the Oschin Schmidt Telescope
on Palomar Mountain and the UK Schmidt Telescope.  The plates were
processed into the present compressed digital form with the permission
of these institutions.} covering the central 38' of the Fornax cluster,
corresponding to our survey area.  The nova candidates are labeled by
number and the galaxy identifications are also indicated.  The positions
of the novae are listed in Table~\ref{tab_fnx_nov_pos}.  Since each
candidate was discovered using slightly different techniques, each will
be discussed in detail below.

\begin{figure}
\plotone{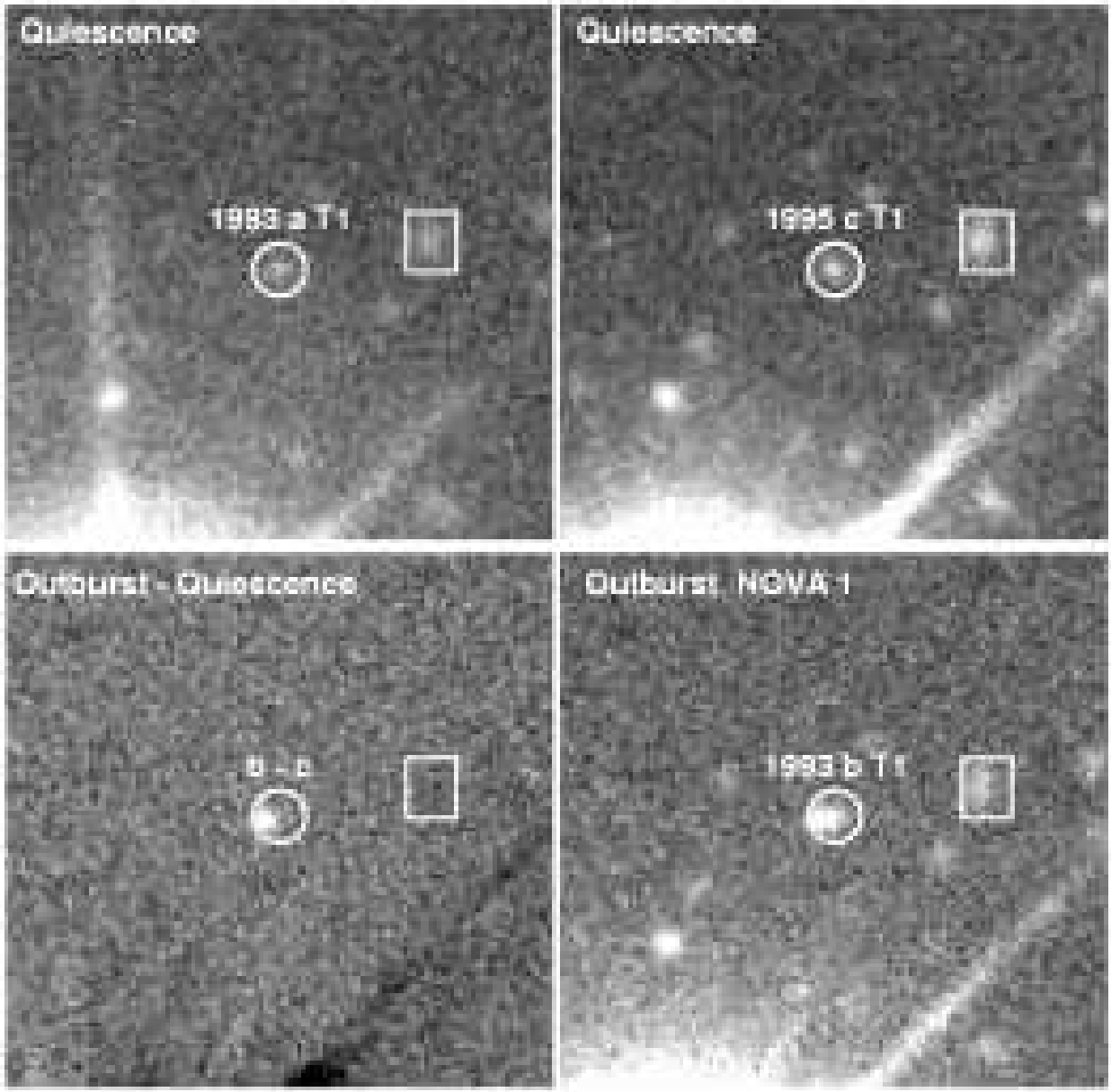}
\caption{Nova 1 discovery images consisting of three T1 filter epochs 
and a subtraction of epoch c from epoch b.  All images are
spatially registered to better than a pixel and the white circle,
indicating the faint star, is in the same location in each image.
For comparison a nearby object is indicated by the white box.
The field-of-view is 44" on a side, North is up and East is to the
left. It is obvious that in outburst the centroid of light within the
circle has shifted.  The subtraction clearly shows that Nova 1 is offset 
from the faint star.
\label{fnx_n1}
}
\end{figure}

{\it Nova 1}.  This candidate was discovered in the T1 image from the 1993
epoch b.  It was easily compared with the T1 image from the 1995 epoch
c which has a deeper frame limit afforded by the longer exposure time
(see Table~\ref{tabforobs}).  We did a careful spatial registration
of the images from other epochs and found that, indeed, this object
is a transient variable meeting the first criterion described above.
It appears near a faint unresolved object, but is at least 10 times
brighter than that object.  Figure~\ref{fnx_n1} shows the three epochs
of T1 images of the region (a, b, and c) and a subtraction of epoch c
from epoch b, which removes the other stars in the field and shows Nova
1 as distinct from the faint point source.

\begin{figure}
\plotone{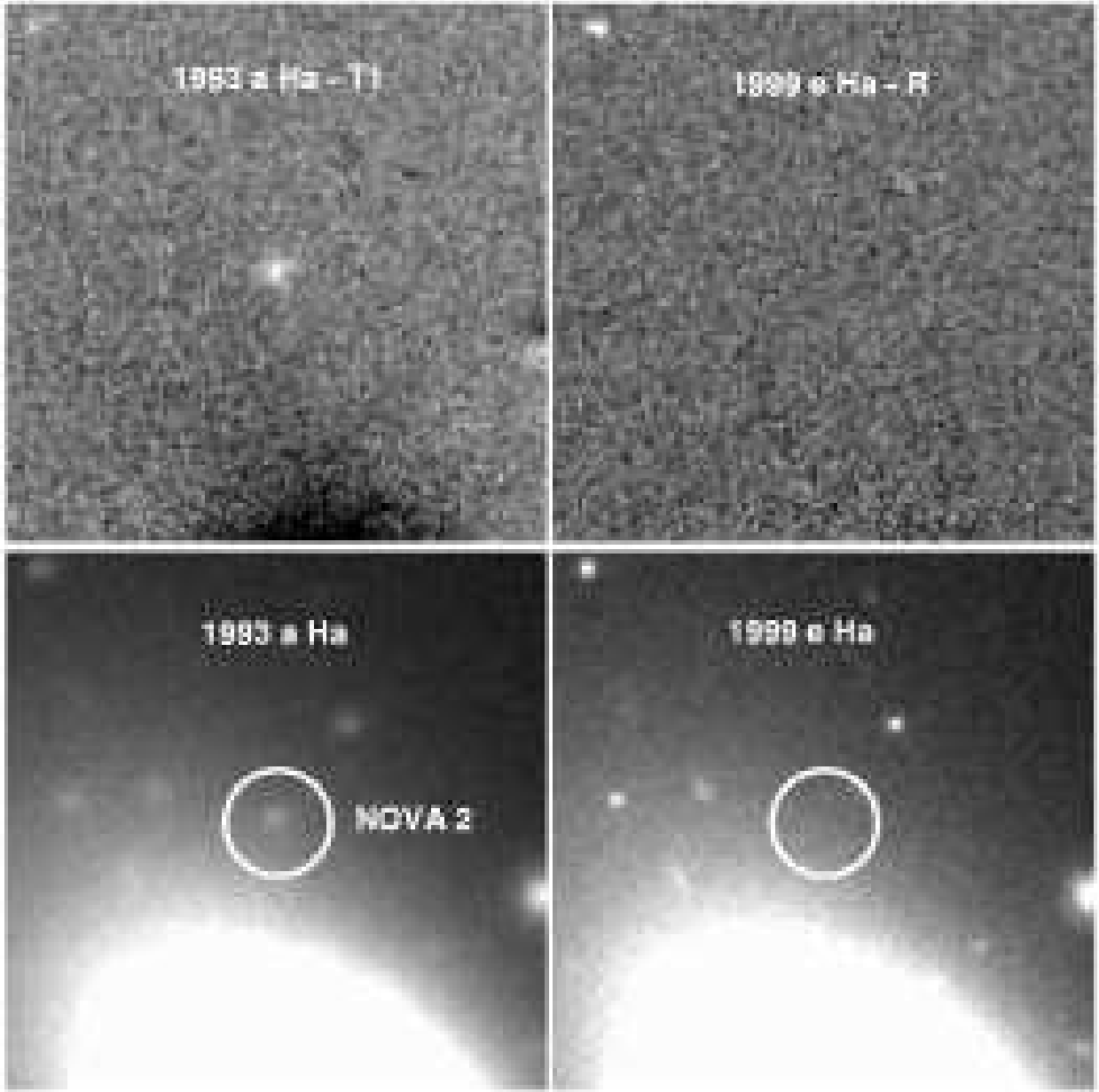}
\caption{Nova 2 discovery images consisting of H$\alpha$ - red continuum 
images from 1993 epoch a and 1999 epoch e (shown in the top two panels) and
the unsubtracted images from the same epochs (shown in the bottom two
panels).  These images illustrate that  Nova 2 is a transient 
H$\alpha$-bright source.  The field-of-view is 27" on a side, North is 
up and East to the left.  The non-transient H$\alpha$-bright source in 
the upper left corner of the images is galaxy B033518.61-355338.0 from the
Automated Plate Measuring Survey \citep{mad90}.
\label{fnx_n2}
}
\end{figure}

{\it Nova 2}.  The 1993 a epoch H$\alpha$ - T1 image revealed an
H$\alpha$-bright source near the galaxy NGC 1389.  Our deep H$\alpha$
image from from the 1999 e epoch allowed us to easily confirm that this
source was transient and an excellent nova candidate, as illustrated
in Figure~\ref{fnx_n2}.  In addition, we confirmed that the object was
not visible in images from any other year.

\begin{figure}
\plotone{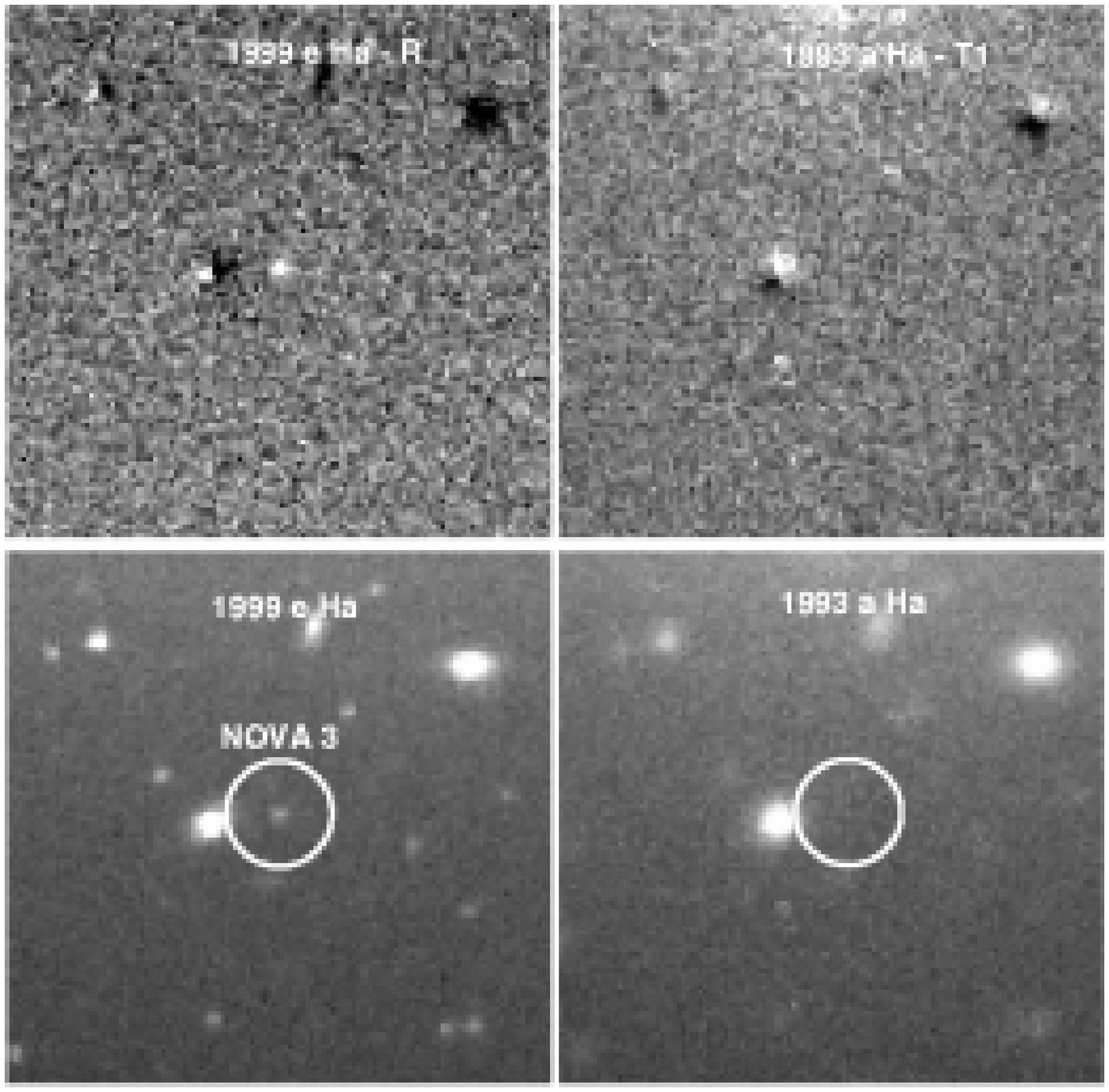}
\caption{Nova 3 discovery images consisting of H$\alpha$ - red continuum
images from the 1999 epoch e and 1993 epoch a (shown in the top two panels)
and the unsubtracted images from the same epochs (shown in the bottom two
panels) demonstrating that Nova 3 is a transient H$\alpha$-bright source.
The field-of-view is 44" on a side, North is up and East to the left.
\label{fnx_n3}
}
\end{figure}

{\it Nova 3}.  In this case we found an H$\alpha$-bright source in the
H$\alpha$ - R image from the 1999 e epoch near the galaxy NGC 1399 as
shown in Figure~\ref{fnx_n3}.  We confirmed its transient nature by the
fact that it was undetectable in all H$\alpha$ and R or T1 filter epochs
from other years.  Because we had 5 closely spaced epochs in 1999, we
were able to generate a light curve in H$\alpha$ and B for this candidate
(see below) and measure its decline rate.

\begin{figure}
\plotone{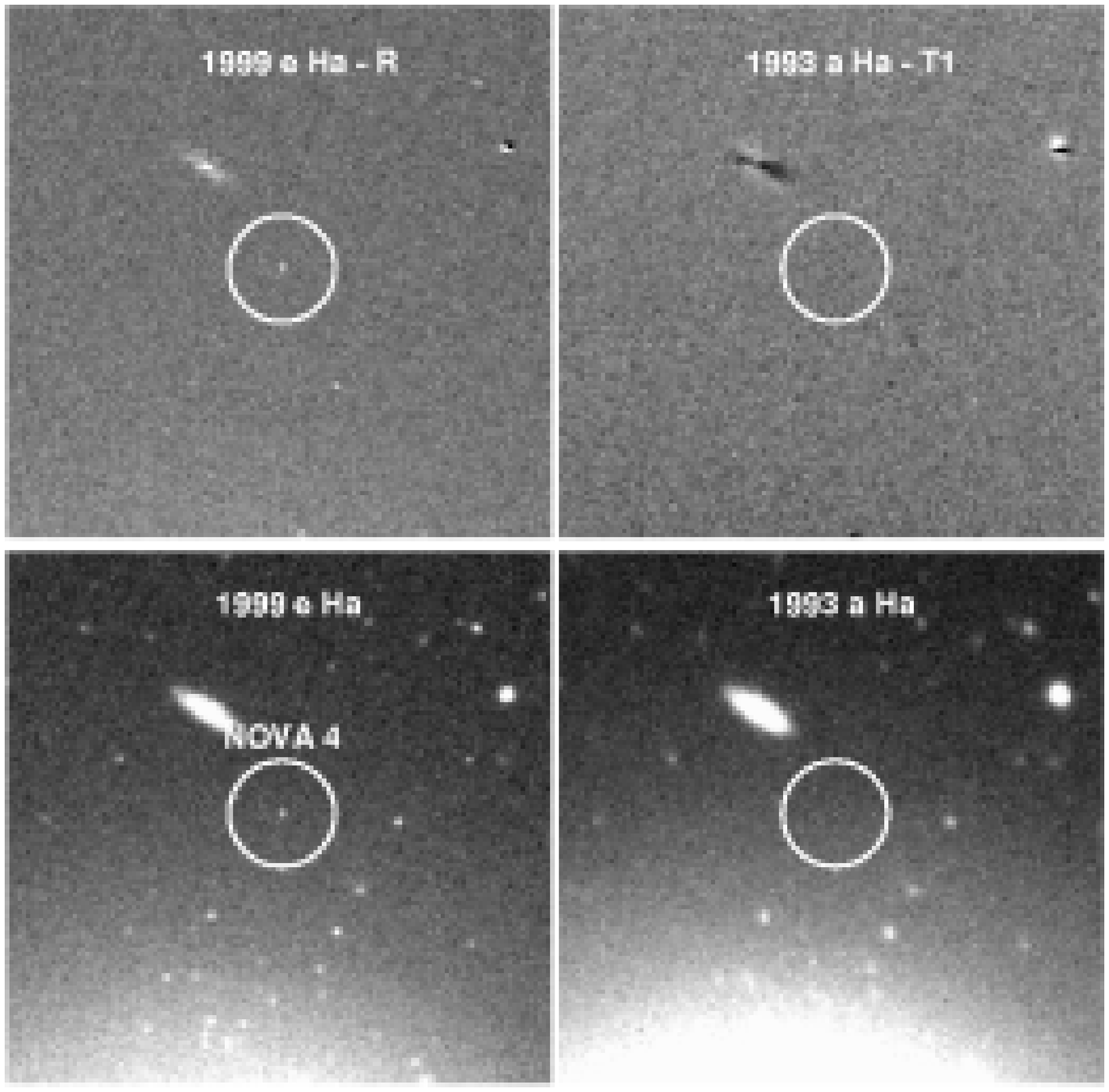}
\caption{Nova 4 discovery images consisting of H$\alpha$ - red continuum
images from the 1999 epoch e and 1993 epoch a (shown in the top two panels)
and the unsubtracted images from the same epochs (shown in the bottom two
panels) demonstrating that Nova 4 is a transient H$\alpha$-bright source.
The field-of-view is 88" on a side, North is up and East to the left.
\label{fnx_n4}
}
\end{figure}

{\it Nova 4}.  This candidate is nearly identical to Nova 3 in the way
it was discovered, although its magnitude is fainter.  It is shown in
Figure~\ref{fnx_n4}.  The light curve is displayed in the next section
and shows an offset between the H$\alpha$ and B-band brightness of 2
magnitudes, similar to that shown by the novae in M31 \citep{cia90a}.
Due to its faintness, we were unable to derive a decline rate in the
B-band for this candidate.  It was also not seen in the images from any
other year.

\begin{figure}
\plotone{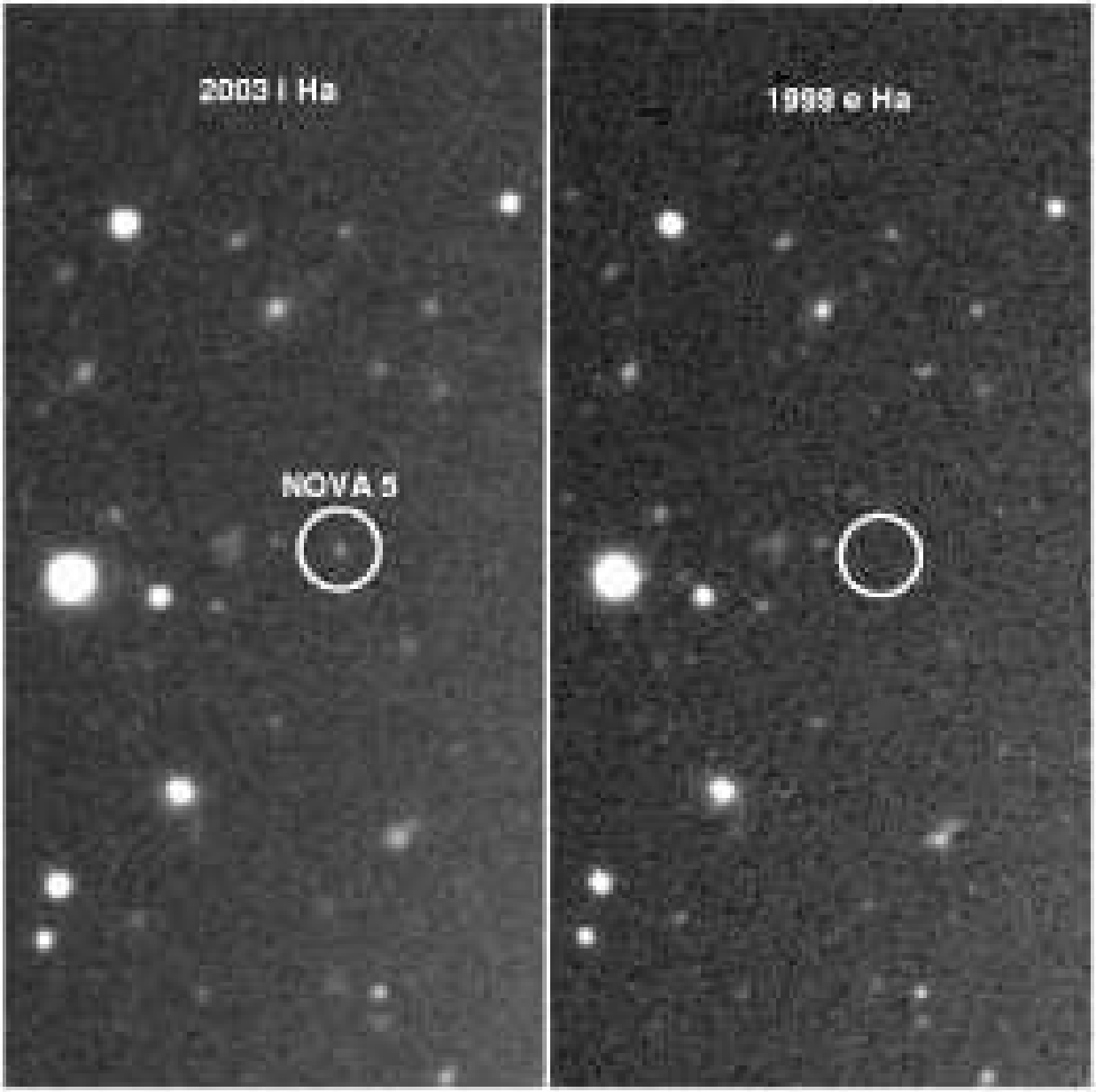}
\caption{Nova 5 discovery images consisting of H$\alpha$ images from the 
2003 i epoch and the 1999 e epoch, showing that Nova 5 is indeed a transient 
H$\alpha$ source.  The fields have North up and East to the left and are 
both 52" wide.
\label{fnx_n5}
}
\end{figure}

{\it Nova 5}.  This candidate was discovered in the 2003 i H$\alpha$
epoch, which did not have a corresponding red continuum image to
allow subtraction.  It was compared with the deeper 1999 e epoch which
proves that the source is transient, as shown in Figure~\ref{fnx_n5}.
It also does not appear on any continuum (B or R or T1-band) image in
any other epoch.

\begin{figure}
\plotone{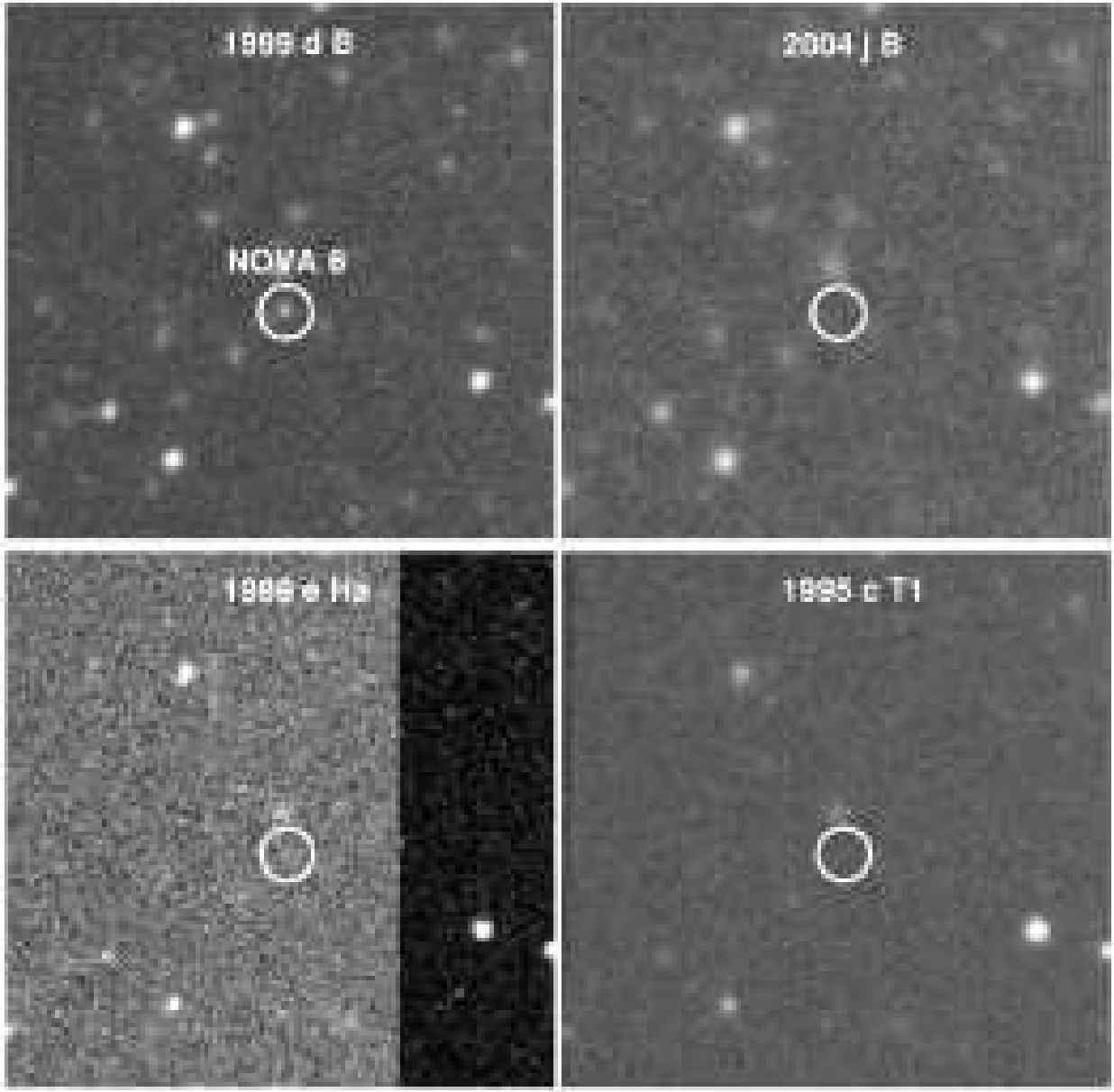}
\caption{Nova 6 discovery images consisting of B-band images from the 
1999 epoch d and the 2004 epoch j, shown in the top two panels.  H$\alpha$
emission can barely be detected in the lower left image from 1999 epoch e
in the part of the chip with a readout problem.  The T1-band image from
1995 epoch c is shown on the lower right to illustrate that this object is
only seen in continuum in the 1999 epochs.  The field-of-view is 55" on a
side, North is up and East to the left.
\label{fnx_n6}
}
\end{figure}

{\it Nova 6}.  We discovered this candidate in the B-band image
from the 1999 epochs after we obtained the 2004 j B-band epoch (see
Figure~\ref{fnx_n6}).  In the 1999 e H$\alpha$ epoch it fell on a part
of the chip that had a readout problem and therefore was missed using
the H$\alpha$ - red continuum subtraction technique.  It was missed in
the comparison between the B-band epochs of a (1993) and e (1999) because
the depth of the 1993 a epoch B-band image was not sufficient for a good
comparison.  After detecting this candidate in the B-band image from 1999,
we then went back and found the corresponding H$\alpha$-bright source in
the H$\alpha$ - R-band image.  We were unable to calibrate this candidate's
H$\alpha$ magnitude to better than 0.5 magnitudes because there were no
calibration sources on the part of the chip with the readout problem.
We were able to generate a B-band light curve (see below) and measure
its decline rate.  The H$\alpha$ magnitude is highly uncertain for this
candidate because of the readout problem mentioned above.

\section{Nova Photometry}

Nova instrumental magnitudes were derived using DAOPHOT \citep{ste87}
aperture photometry.  To account for variable seeing, we set the
aperture radii to 1/2 the FWHM of the stellar profile measured from high
signal-to-noise, well-isolated stars.

Photometric calibration was achieved using a variety of secondary
standards.  For B-band calibration we used the Yale/San Juan Southern
Proper Motion Program \citep{pla98} which had 7 unsaturated stars
in common with our B-band observations in 1999.  This produced a
calibration accurate to 0.1 magnitudes.  For the calibration of the
T1, and R-band photometry we used the R-band magnitudes for the
globular clusters published by \citet{kis99}.  For the calibration we
used measurement apertures that were at least as large as those used
to produce the published magnitudes \citep{gri94}.  Since the T1 and
the R filter cover a similar wavelength range, we simply forced the
T1 magnitudes to the R system.  This calibration is also accurate to 0.1
magnitudes.

For the H$\alpha$ magnitudes, we had no calibrated H$\alpha$ sources in our
fields and no observations of flux standards in our H$\alpha$ filters.
We decided, instead, to use the R-band magnitudes from \citet{kis99} to 
calibrate our H$\alpha$ magnitudes.  This procedure is less than ideal
because it introduces several uncertainties.  Novae are H$\alpha$ emission
line objects and globular clusters are continuum objects.  This has been
overcome in the past by using a filter that is broad enough to include 
most of the H$\alpha$ light and then assuming the H$\alpha$ light fills
the bandpass, i.e. by just quoting a filter magnitude and not a line flux
(see e.g. Ciardullo et al. 1989).  This will hold for the 1993 a epoch
which used an H$\alpha$ filter that is correctly redshifted for Fornax and
a width that includes most of the H$\alpha$ light (nova expansion
velocities up to 1700 km s$^{-1}$).  For the other epochs which used an
unshifted H$\alpha$ filter, most of the H$\alpha$ light will fall toward
the red edge of the filter bandpass where the transmission is dropping
rapidly.  Without spectra of each nova, the amount of flux lost is unknown.
For objects from these filters, there will be an unknown systematic error
in the filter magnitudes.  This also makes calculating the limiting
magnitudes for these epochs uncertain as well.

With this in mind,
we simply calculated the offset between the R system and our instrumental
H$\alpha$ magnitudes using the globular clusters, and applied this
offset to force our instrumental H$\alpha$ magnitudes to the R system.
The number of objects used to perform the calibration varied from 13
to 30.  For the 1993 a epoch the external error in H$\alpha$ is better than 0.2
magnitudes, while for the other epochs the external H$\alpha$ error is such
that the magnitudes quoted are overestimates (the novae are probably brighter).

\begin{figure}
\includegraphics[width=0.95\linewidth,bb= -20 60 550 730,clip=true]{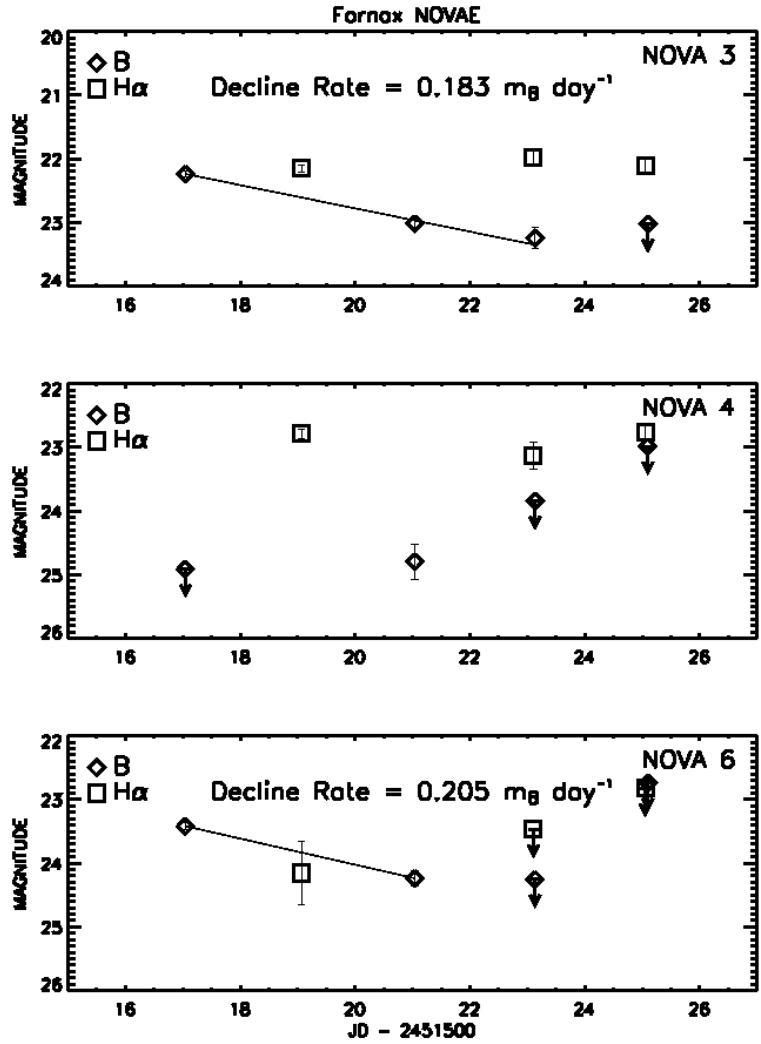}
\caption{Light curves for Fornax Nova
3, Nova 4, and Nova 6.  The squares are the H$\alpha$ magnitudes and the
diamonds are the B-band magnitudes.  Points with downward pointing arrows
are lower limits on the magnitudes.  The decline rates were measured for
Nova 3 and Nova 6 by using error-weighted linear fits, indicated by the 
thin solid lines.
\label{fnx_lcs}
}
\end{figure}

Table~\ref{tab_fnx_nov_phot} lists the calibrated magnitudes
for each nova and Figure~\ref{fnx_lcs} shows the B and H$\alpha$ light
curves for the multiply observed novae (Nova 3, Nova 4, and Nova 6).
We used error-weighted linear fits to derive a decline rate for Nova 3 of
0.183 m$_B$ day$^{-1}$ and for Nova 6 of 0.205 m$_B$ day$^{-1}$.  The fits
are are indicated in Figure~\ref{fnx_lcs} by
the thin solid lines.  The errors quoted in Table~\ref{tab_fnx_nov_phot}
and plotted in Figure~\ref{fnx_lcs} are the internal measurement errors
and in most cases are smaller than the external calibration errors.

\section{Are These Variables Really Novae? \label{novae_verify}}

Could these six candidates be some other kind of variable and not novae?
Without spectra, it is impossible to be certain.  All of our candidates
are transient according to the criterion discussed in the first paragraph
of \S\ref{novae_det}, thus making periodic or semi-regular variables
unlikely contaminators.  Four of the six candidates (Novae 2, 3, 4, and 6)
are H$\alpha$-bright and transient, virtually ruling out other possible
variables (see below).  Other types of variables that are transient
include background AGN, $\gamma$-ray bursts (GRBs), supernovae (SNe),
and microlensing events.  We must resort to statistical arguments in an
attempt to eliminate each of these classifications for Novae 1 and 5.

The long-term quasar variability study of \citet{hel01} shows that 
the rms variability of their sample is 0.45 mag in the blue and 0.33
mag in the red (see their Figure~4).  Both Nova 1 and Nova 5 are measured
at least a magnitude above the plate limit in the red (T1 for Nova 1 and
H$\alpha$ for Nova 5) making it unlikely that these could be background AGN
that have varied on timescales of a year.  We also checked the positions of
all our novae against the deep {\it Chandra} Fornax X-ray survey of 
\citet{sch04} and 
found no coincidences with any X-ray sources.  As an additional check we
consulted the NASA/IPAC Extragalactic Database (NED) and found no
coincidences with Nova 1 or Nova 5 out to a search radius of 30 arcseconds.

The total GRB rate derived from {\it BATSE} detections is about 600
yr$^{-1}$, of which $\sim$40\% are 'dark', or not visible in the optical
\citep{zha04}.  This gives an optical GRB rate of close to 1 per day
over the entire sky.  Our survey was roughly 30 arcminutes on a side or
0.25 square degrees, so the a priori probability of detecting a GRB in 
our survey field in any given epoch is $6.1\times10^{-6}$.

An estimate of the SN rate in the local universe is given by \citet{cap01}
who gives the rate for all types of SNe of $1.21\pm0.36\ h^2$ SN
$10^{-10}L_{\odot,B}\ 100$ yr$^{-1}$.  If we compare this to the nova
rate from \citet{cia90b} of 24 novae $10^{-10}L_{\odot,B}$ yr$^{-1}$
we see that novae are $4.0\times10^3$ times more common than SNe in the
local universe (using H$_0 = 70$ \kms\ Mpc$^{-1}$).  This makes it more
likely that our candidates are novae rather than SNe in the Fornax cluster
that declined substantially before we detected them.  It is also very
unlikely that a SN in the Fornax cluster would go undetected since it would
reach a brightness of 12th magnitude and would decay with a timescale of
months.

That being said, we can see SNe 100 times more distant than novae (SN reach
M$_B = -19$).  We must, therefore, consider SNe in undetected background
galaxies and intracluster SNe in background clusters.  \citet{sch04} did
find a background cluster in their deep X-ray survey of Fornax, but it is
located 2.6 arcminutes from Nova 1 and has a diameter of 1.4 arcminutes in
the soft band and so it is not likely that Nova 1 is a SN associated with
this cluster.  We cannot rule out the possibility that the object next to
Nova 1 is an unresolved galaxy and that Nova 1 is a SN in that galaxy.
A spectrum of this object would resolve this issue, but it would require 
a very large telescope since it is very faint (R $\sim$ 24.5).

Our deepest B-band image has a plate limit of 26.  No objects are found to 
this limit at the positions of any of the novae (with the exception of Nova
1).  If Nova 5 was really a SN in an undetected background galaxy or
cluster it would have to be fainter than B of 26 and therefore much fainter
than the SN itself.  Although unlikely, we cannot rule this possibility out.

To assess the contamination from microlensing events, we will turn to a
microlensing survey of M87 in Virgo by \citet{bal04}.  Using the WFPC2
camera on {\it HST} every day for a month, this study detected 1 strong
microlensing candidate.  It's peak magnitude was fainter than 24 and would
have been barely detected in our deepest survey epochs.
They report that this detection rate is consistent with the sensitivity 
and coverage of their survey and their microlensing models that
included self-lensing from stars in M87, lensing from the M87 halo, the
Virgo halo, and foreground lensing from the Milky Way \citep{bal04}.

Our sensitivity to microlensing events would be considerably less than
their survey due to our lower resolution and shallower depth (our faintest
measurement in the B-band is 24.8, their faint limit is 28.5 assuming a
3$\sigma$ detection, see Baltz et al. 2004).  The microlensing models
from \citet{bal04} show that self-lensing and lensing from the Milky Way
halo provide the smallest contributions to the microlensing rate.  The 
largest microlensing rates come from 
the Virgo halo, however, the rate depends strongly on the concentration of
lenses toward the center of M87 where there are more sources to be lensed.
This means that our much larger survey area does not drastically increase
our sensitivity to microlensing events relative to theirs. This is because even
though we include many more lenses (the Fornax cluster halo), we only
include a factor of a few more sources to be lensed (the stars of NGC
1399 and the other three galaxies) which would not make up for the much
brighter faint limit and the lower resolution of our survey.

From the models of \citet{bal04}, we assume that the probability for a 
microlensing event drops with underlying stellar density.  From their 
observed rate and the relative similarity of
M87 and NGC 1399, we would expect to barely detect one event in NGC 1399 in
one month.  We can eliminate the microlensing classification for our
H$\alpha$-bright candidates because microlensing events are not
H$\alpha$-bright.
The two remaining nova candidates, Nova 1 and Nova 5, are distant from 
background stellar light and so the probability of detecting a lensing 
event in those regions should be much less than predicted in the study 
by \citet{bal04}.

Admitting that the classification for all our candidates is not 
iron-clad, these arguments indicate the most probable interpretation is
that our candidates are novae.  We will assume so for our subsequent 
analysis.

\section{Nova Properties}

We were fortunate that three of the novae erupted during the five epoch
run in 1999.  The two measured decline rates (see Figure~\ref{fnx_lcs})
of 0.183 m$_B$ day$^{-1}$ for Nova 3 and 0.205 m$_B$ day$^{-1}$ for
Nova 6 place both in the fast nova category.  If we assume a distance
modulus to the Fornax cluster of $(M~-~m)_0 = 31.35$ \citep{kis99}, then
our brightest B magnitude for Nova 3 is at M$_B = -$9.2.  Typically,
novae reach peak B magnitudes of M$_B = -8.0$, so Nova 3 is clearly
a brighter-than-average nova (see, e.g., Figure~5.3 in Warner 1995).
The flat shape of the H$\alpha$ light curve indicates that we have caught
this nova near H$\alpha$ maximum and during the earliest phases of the
B decline.  The brightness of Nova 3 is consistent with its fast decline
rate since decline rate is known to correlate with nova peak brightness
\citep{sha81}.  Given the decline rate of Nova 6, it is possible that
this nova was at least as bright as Nova 3 at maximum, but that we are
observing it in its later decline phase.  The large uncertainty in the
H$\alpha$ magnitude for this nova does not allow us to compare its B and
H$\alpha$ magnitudes.  

With the uncertainties in the survey H$\alpha$ limit mentioned above it is
difficult to quantitatively assess the nova populations in the cluster.
As stated in \S\ref{fnx_intro}, we predicted 12 novae to have visible 
H$\alpha$ emission down to the limit of M$_{H\alpha} = -7.5$ in any 
given H$\alpha$ epoch, yet we detected only 5 novae in our 3 H$\alpha$ 
epochs, a factor of 7 less than we expected.  There are several factors
that could reduce the number of novae we detected.  Intracluster novae with
velocities that would shift their H$\alpha$ emission even farther to the
red would lose a significant amount of their flux in the epochs using the
unshifted H$\alpha$ filter (Epochs e, g, h, and i).  Accounting for the
velocity dispersion of the intracluster population and the individual
galaxies, it is possible that we are down by at least a factor of 4 
by using the unshifted H$\alpha$ filter for two of the three years of the
survey.

We point out that our earlier results indicate that novae are predominantly 
associated with the bulge population of M81 \citep{nei04}.  Novae,
therefore, have the potential to provide some clues to the source of the 
intracluster light.  A long-term, densely time-sampled study is required 
to properly assess the nova parent population in Fornax.  Since all 
regions in the survey were sampled with equal frequency, an analysis of the
spatial distribution of the Fornax novae is worthwhile, though it cannot
be conclusive.

\section{The Spatial Distribution of the Novae}

We deliberately limited ourselves to two simple questions about the
spatial distribution of novae in the Fornax cluster.  First, assuming
that novae follow the bulge light in galaxies \citep{nei04}, how many
novae did we miss in the centers of the galaxies?  Second, based on the
light profiles of the galaxies, how many novae are associated with the
cluster potential and not with any galaxy?

To answer these questions, we required the light profiles of the four
galaxies in our survey region: NGC 1399, NGC 1404, NGC 1389, and NGC 1387.
Since NGC 1399 has an extended halo, we used the study of \citet{sag00}
for its light profile.  They used a combination of wide-field photographic
plates and narrower-field CCD data to generate a profile that extends
nearly 2000 arcseconds from the center of NGC 1399.  For the other
profiles, we used our B-band mosaic image from the 1999 epochs.
We measured their profiles using the ellipse task in the STSDAS {\bf
isophote} package.  Isophotes were fit on background subtracted images
with the background being determined by the region in the chip most
distant from the galaxy.  The task was given the initial center position,
position angle (PA), and ellipticity and allowed to vary these parameters
at each semi-major axis position, which was set to grow geometrically
and not linearly.  This allowed the outer isophotes to contain enough
signal to produce reasonable fits.

To answer the incompleteness question, we used artificial stars to
determine our detection limit for each galaxy.  Our goal was to produce a
simple correction to the number of novae found in the galaxies.  We used
the artificial stars to ask the question, how close to the nucleus of
the galaxy could the nova that was discovered in that galaxy be found?
By comparing this radial limit to the light profile, we know how much
light in the galaxy was effectively unsurveyed for novae of the brightness
already discovered, and can thus make a correction based on the fraction
of the total light unsurveyed.

DAOPHOT was used to generate a point spread function (PSF) and then used
to add artificial stars around each galaxy.  The galaxy images were then
median-filtered, and subtracted, and (as described above) blinked to
see where the artificial novae become undetectable.  In all cases this
limit was determined by the residual light left by the subtraction of
the median-filtered image.  Thus, the inner profiles of the galaxies
strongly affected this detection limit.

\begin{figure}
\includegraphics[angle=90,width=\linewidth]{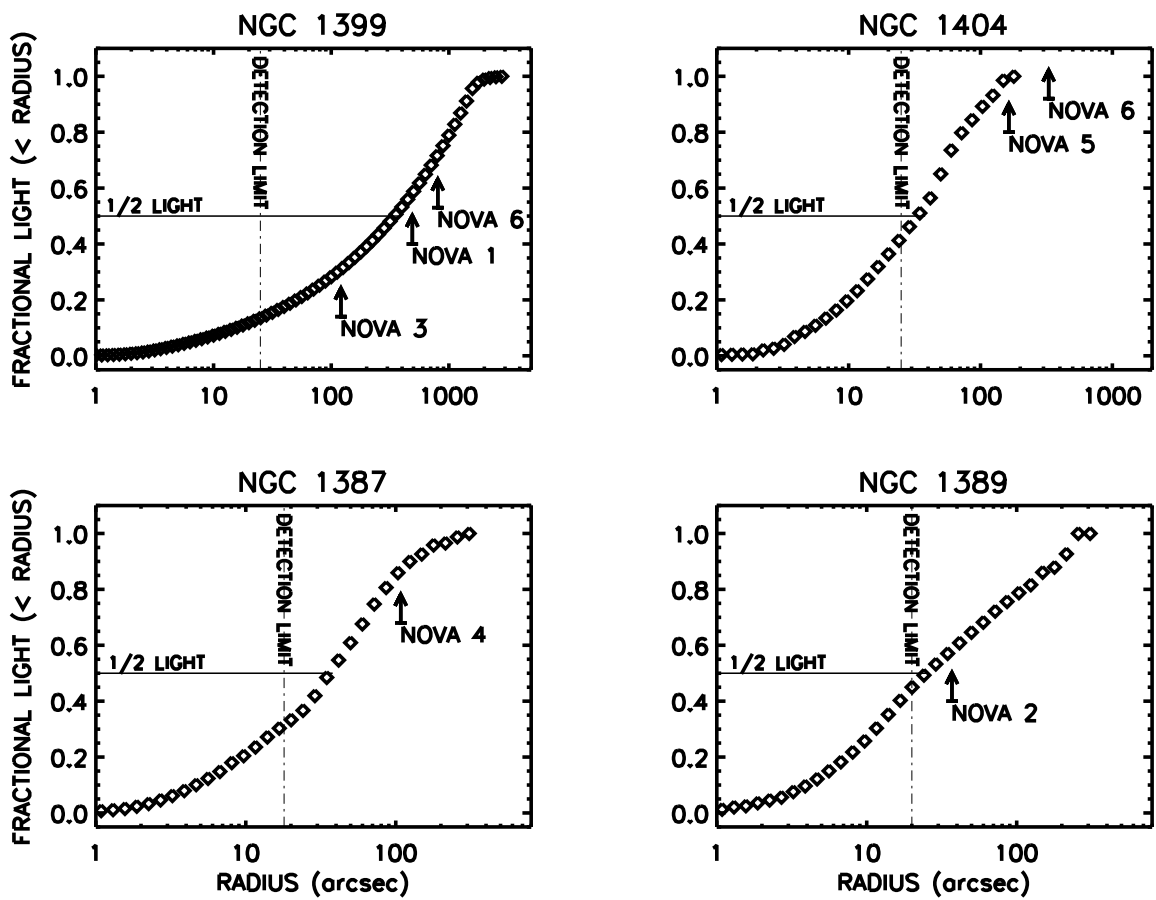}
\caption{Cumulative light profiles
for NGC 1399, NGC 1404, NGC 1387, and NGC 1389 in fraction of total light
versus radius in arcseconds indicated by the diamonds.  The detection
limits are indicated by the dot-dashed line and the positions of the novae
are indicated by the upward pointing arrows and labeled by the nova id.
The 1/2 light level for each galaxy is labeled and indicated with a thin 
solid horizontal line.
\label{fnx_profs}
}
\end{figure}

Figure~\ref{fnx_profs}  shows the cumulative light profiles for the four
galaxies, normalized to the total light.  We set the total light limit
at 26 B magnitudes per square arcsecond for the galaxies we measured.
The profile of NGC 1404 was contaminated by a very bright star 180"
from the galaxy, so we set a limit at that distance of 25 B magnitudes
per square arcsecond.

The first thing we notice is the extent of the halo of NGC 1399.
Nearly the entire survey area is enclosed by this halo.  We will
assume that this halo actually traces the inner potential well of the
cluster itself as suggested by \citet{mer84}.  We also see that Nova
2 is clearly associated with NGC 1389: it is close to the half-light
radius, but the association of the other novae is not straightforward.
Nova 4 appears at a radius that encloses over 80\% of the light of
NGC 1387, while Nova 5 and Nova 6 are even farther outside of NGC 1404.
Interpreting the nova placement in NGC 1399 is complicated by the extended
halo, although it seems clear that Nova 3 belongs with that galaxy.
The extent of the halo of NGC 1399 means that Nova 1 is located just
beyond the half-light radius, but clearly it is more properly placed in
the cluster rather than in that galaxy (see Figure~\ref{fnx_nov_pos}).
\citet{gri94} found that the transition from being bound to NGC 1399 and
being bound solely to the cluster occurs at a distances between 1\minpnt5
and 5\minpnt5 from the center of NGC 1399 (see their Figure 2).  Nova 1 is
8\minpnt2 from the center of NGC 1399 and well past this transition.
We also placed Nova 6 on the NGC 1399 light
profile and it appears at a radius that encloses only 65\% of the light.
To be conservative, we'll assume that 4 of our novae (2, 3, 4, and 5)
are associated with the galaxies in the cluster and 2 (1 and 6) are not.
But how many novae did we miss in the centers of the galaxies?

To generate a correction for missed novae, we look at the fraction of
light unsurveyed, as determined by our detection limits and plotted as
the vertical dot-dashed lines in Figure~\ref{fnx_profs}. For NGC 1399
this fraction of unsurveyed light is 0.15, for NGC 1404 it is 0.4, for
NGC 1387 it is 0.3 and for NGC 1389 it is 0.45, giving a total unsurveyed
light fraction of $1.3/4 = 0.325$.  Since we found, on average, 1 nova
associated with a galaxy, this would essentially add another $0.325\times4
= 1.3$ novae to our number associated with galaxies, giving a final
score of 5.3 novae in and 2 novae outside of galaxies in the cluster.

\section{A Comparison with the X-Ray Gas}

\begin{figure}
\includegraphics[width=\linewidth]{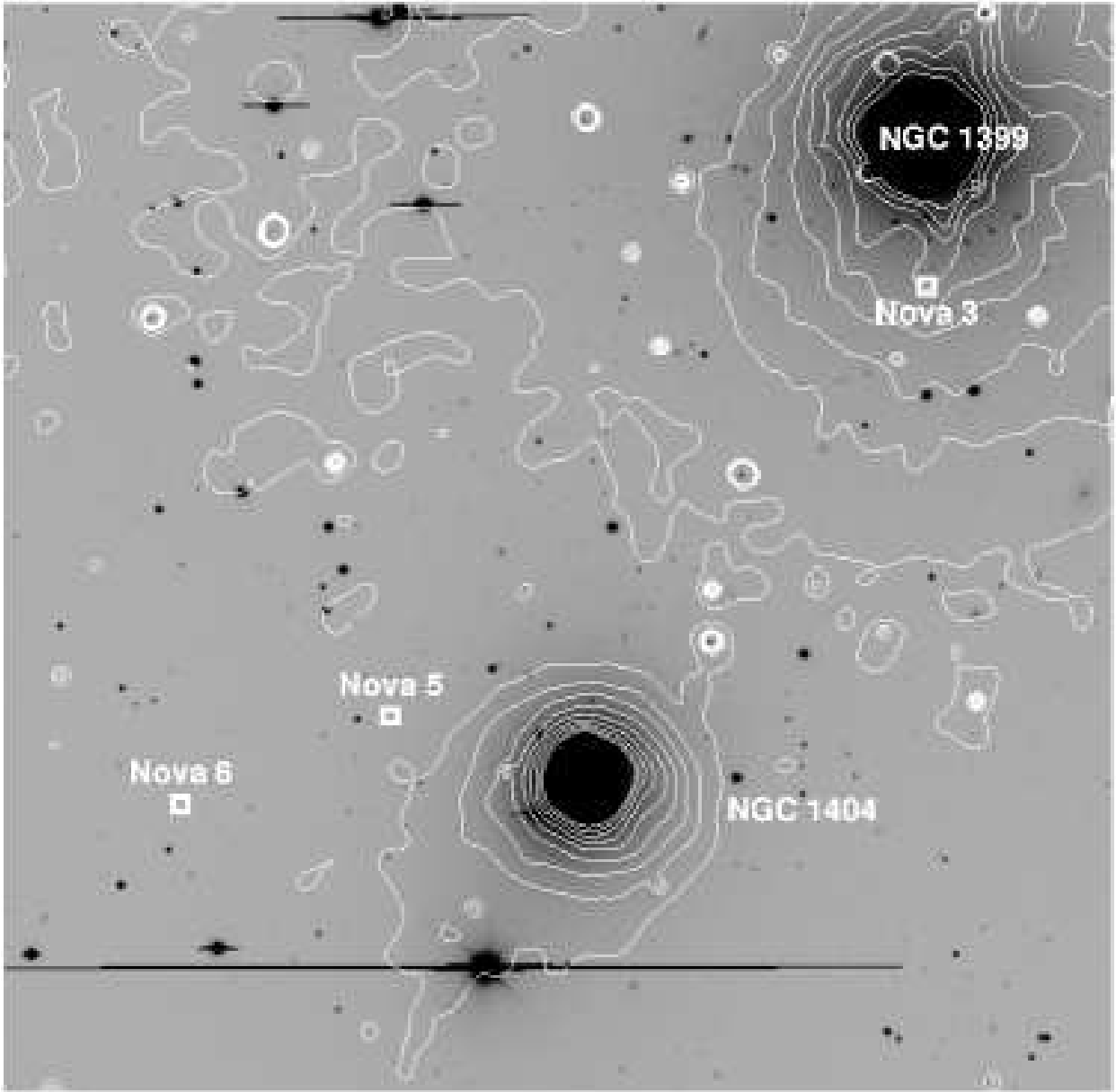}
\caption{X-ray contours overplotted on the B-band image from epoch j (2004).
Nova 3 is coincident with the jet on the south side of NGC 1399.  NGC 1404
appears to have a tail in the X-ray gas that indicates motion toward NGC
1399 \citep{sch04}.  Nova 5 and Nova 6 are outside of the bow-shock region 
and therefore
are not likely to be gravitationally bound to NGC 1404.  The X-ray contours
are from the energy range of 0.3 to 1.5 KeV.  The lowest contour
is at a flux of 7.7$\times$10$^{-17}$ erg cm$^{-2}$ s$^{-1}$
arcmin$^{-2}$ and the highest contour is at a flux of 4.9$\times$10$^{-16}$
erg cm$^{-2}$ s$^{-1}$ arcmin$^{-2}$.
\label{fnx_xray}
}
\end{figure}

The X-ray study of the Fornax cluster by \citet{sch04} offers some
interesting evidence concerning the association of the novae with the
galaxies nearby.  Figure~\ref{fnx_xray} shows the X-ray flux levels
near NGC 1399 and NGC 1404 generated from a soft band image (0.3 - 1.5
KeV) with an exposure time of $\sim$45 ksec \citep{sch04}.  Ten evenly
spaced contour levels ranging from a flux of 7.7$\times$10$^{-17}$
erg cm$^{-2}$ s$^{-1}$ arcmin$^{-2}$ to a flux of 4.9$\times$10$^{-16}$
erg cm$^{-2}$ s$^{-1}$ arcmin$^{-2}$ are overplotted on our B-band image
from epoch j (2004).  Three of our novae (Nova 3, Nova 5 and Nova 6)
fall in this region and are indicated by small boxes and labeled with
their identifications.  As pointed out in \citet{sch04}, NGC 1404 has a
bow-shock on the side facing NGC 1399 due to its motion toward NGC 1399.
The comet-like tail they describe is apparent in the lowest contour
of Figure~\ref{fnx_xray}.  This figure
solidifies the association of Nova 6 with intracluster space and not
with NGC 1404 because it is well outside the lowest X-ray contours for
NGC 1404 and well outside the bow-shock of the X-ray gas in NGC 1404.

Using our optical light profiles alone (see Figure~\ref{fnx_profs}), we
would associate Nova 5 with NGC 1404.  We see from Figure~\ref{fnx_xray}
that Nova 5 is outside of the bow-shock of NGC 1404.  Without a detailed
investigation of the dynamics of NGC 1404, we cannot conclusively say that
the stars outside the bow-shock are not bound to NGC 1404, but the
placement of Nova 5 is suggestive of its membership in the intracluster
population.  This evidence suggests that we might revise our score to 
4.3 novae in galaxies and 3 novae in intracluster space.

\section{The Intra-Cluster Light in Fornax}

Assuming that novae trace the intracluster light in a galaxy cluster, we
calculate that the ICL contributes $3/( 4.3 + 3 ) = 41$\% of the total
light in the Fornax cluster.  This is very close to the estimate of
40\% by \citet{the97}, who used intergalactic planetary nebulae (PNe).
This is likely an overestimate of the ICL due to contamination by background
Ly$\alpha$-emitting galaxies masquerading as PN \citep{cia02}.  If we move
one nova from the out to the in category (e.g., Nova 5), we get an ICL 
fraction of $2/( 5.3 + 2 ) = 27$\% of the total light, similar to the 
ICL fraction of 22\% estimated for Virgo by \citet{fel98} from 
intergalactic PNe.  If we also assume that one of our intracluster novae 
is really a background SN (e.g., Nova 1), then we get an ICL fraction of 
$1/(5.3 + 1) = 16$\%.  All these numbers are within the range of other 
measures of ICL.  {\it Our key result is that we have demonstrated for 
the first time that intergalactic tramp novae very likely exist in the 
Fornax cluster, and that they can be useful tracers of the ICL}.

\section{Future Work}

Clearly, a longer-term, densely time-sampled survey in H$\alpha$ and
continuum light of this region would allow us to provide a much more
accurate measure of the ICL in Fornax.  Using an H$\alpha$ filter that
accounts for both the redshift of Fornax and the velocity dispersion of the
cluster and member galaxies as well as the range of nova line widths would improve
the number of novae detected in each epoch by a factor of up to 7 over
the survey reported here.  This kind of survey would not
only bolster the statistics by adding more novae, but would also provide
light curves which can be used to better verify novae and estimate their
peak brightnesses.  It would also allow us to estimate the nova rates in
each galaxy and in the ICL.  This would be the first step toward assessing
the source population of the ICL and estimating
the effect of the galaxy cluster environment on close binary stars.

\section{Conclusions}

1. Intra-cluster novae do exist in the Fornax cluster and their
distribution is consistent with $\sim$16 - 41\% of the total light in
the cluster being contributed by the ICL.  The high end of this range
is close to the estimate of the ICL in Fornax using intracluster PNe
by \citet{the97}.  The low end is close to the estimate of the ICL in
Virgo using intracluster PNe by \citet{fel98}.

2. Tramp novae are very interesting and useful as tracers of the ICL.
Their brightness
at peak (M$_B$ up to $-$10) allows them to be visible out to the
Coma cluster with current telescopes.  New objects can be accumulated with
each observation epoch, increasing the accuracy of the measure of the ICL
with time.  As our understanding of nova parent populations improves, they 
may become useful for constraining the origin of the tidal debris in clusters.



\acknowledgments

We wish to thank Martin Bureau, Jeremy Mould, Tim Abbott, Knut Olsen, and 
Chris Smith for their support in acquiring Fornax cluster survey epochs in 
2003 and 2004.  We would also like to thank the anonymous referee for very
useful comments on the manuscript.  This research has made use of the 
NASA/IPAC Extragalactic
Database (NED) which is operated by the Jet Propulsion Laboratory,
California Institute of Technology, under contract with the National
Aeronautics and Space Administration.




\clearpage



\clearpage



%
%
\begin{deluxetable}{clllllc}
\tabletypesize{\small}
\tablecaption{Fornax Cluster Observations \label{tabforobs}}
\tablewidth{0pt}
\tablehead{ \colhead{Epoch ID} &
        \colhead{Date} & \colhead{Telescope} & \colhead{Detector} &
        \colhead{Filter} & \colhead{Exptime} & \colhead{Seeing} \\
\colhead{} & \colhead{} & \colhead{} & \colhead{} & \colhead{} &
	\colhead{(s)} & \colhead{(")}\\
} 
\startdata 
a & NOV-1993	& 4m-CTIO	& Tek2048	& B	&	900 & 1.5 - 2.2\\
  &\ \ \ \ \ \ \ " &\ \ \ \ "      &\ \ \ \ \ \ "  & T1	&	900 & 1.4 - 1.7\\
  &\ \ \ \ \ \ \ " &\ \ \ \ "	&\ \ \ \ \ \ "  & H$\alpha$\tablenotemark{a} &	4500 & 1.4 - 1.9\\
\hline
b & DEC-1993	& 1.5m-CTIO	& Tek2048	& T1	&	5400 & 1.1 - 1.3\\
\hline
c & OCT-1995	& 1.5m-CTIO	& Tek2048	& T1	&	9000 & 1.1 - 1.5\\
\hline
d & 05-DEC-1999	& 4m-CTIO	& Mosaic 8K	& B	&	4200 & 1.0\\
e & 07-DEC-1999	&\ \ \ \ "	&\ \ \ \ \ \ "	& R	&	1260 & 0.9\\
  &\ \ \ \ \ \ \ " &\ \ \ \ "	&\ \ \ \ \ \ "	& H$\alpha$\tablenotemark{b} &	7200 & 0.8\\
f & 08-DEC-1999	&\ \ \ \ "	&\ \ \ \ \ \ "	& B	&	4200 & 1.1\\
g & 11-DEC-1999	&\ \ \ \ "	&\ \ \ \ \ \ "  & R	&	1260 & 1.2\\
  &\ \ \ \ \ \ \ " &\ \ \ \ "	&\ \ \ \ \ \ "  & H$\alpha$\tablenotemark{b} &	1800 & 1.2\\
  &\ \ \ \ \ \ \ " &\ \ \ \ "	&\ \ \ \ \ \ "	& B	&	1800 & 1.4\\
h & 12-DEC-1999	&\ \ \ \ "      &\ \ \ \ \ \ "  & R	&       1260 & 1.6\\
  &\ \ \ \ \ \ \ " &\ \ \ \ "	&\ \ \ \ \ \ "	& H$\alpha$\tablenotemark{b} &	3600 & 1.4\\
  &\ \ \ \ \ \ \ " &\ \ \ \ "	&\ \ \ \ \ \ "	& B	&	3000 & 1.2\\
\hline
i & 27-FEB-2003	& 4m-CTIO	& Mosaic 8K	& H$\alpha$\tablenotemark{b} &	6000 & 1.2\\
\hline
j & 14-JAN-2004	& 4m-CTIO	& Mosaic 8K	& B	&	8100 & 1.8
\enddata
\tablenotetext{a}{6600/75\AA}
\tablenotetext{b}{6563/80\AA}
\end{deluxetable}

%
%
\begin{deluxetable}{rcccrr}
\tablecaption{Fornax Cluster Nova Positions\label{tab_fnx_nov_pos}}
\tablewidth{0pt}
\tablehead{
\colhead{} & \multicolumn{2}{c}{Position}\\
\colhead{} & \multicolumn{2}{c}{(J2000)}\\
\colhead{} & \multicolumn{2}{c}{\rule[15pt]{105pt}{0.5pt}} & & &
	\colhead{Distance}\\
\colhead{Nova} & \colhead{RA} & \colhead{Dec} &
\colhead{Epoch(s)} & \colhead{Galaxy} & \colhead{(arcsec)} }

\startdata
1       &03:39:06.2 &-35:23:48	&b      &NGC 1399	&489\\
2       &03:37:12.1 &-35:44:11	&a      &NGC 1389	&37\\
3       &03:38:29.0 &-35:29:02	&d-h    &NGC 1399	&121\\
4       &03:36:55.8 &-35:28:39	&d-h    &NGC 1387	&108\\
5       &03:39:04.5 &-35:34:45	&i      &NGC 1404	&164\\
6       &03:39:18.4 &-35:35:55	&d-h    &NGC 1404	&329
\enddata
\end{deluxetable}

%
%
\begin{deluxetable}{rrclrc}
\tabletypesize{\footnotesize}
\tablecaption{Fornax Cluster Nova Photometry\label{tab_fnx_nov_phot}}
\tablewidth{0pt}
\tablehead{
	       & \colhead{J. D.}\\
\colhead{Nova} & \colhead{(+240000)} & \colhead{Epoch} & \colhead{Filter} 
	& \colhead{m} & \colhead{Err(m)}
}
\startdata
1       &49331.53 & b & T1\tablenotemark{a}	& 21.9	& 0.1\\
\\
2	&49299.84 & a & H$\alpha$ & 21.1 & 0.4\\
\\
3	&51519.07 & e & H$\alpha$ & 22.1 & 0.1\\
	&51523.10 & g & H$\alpha$ & 22.0 & 0.1\\
	&51525.05 & h & H$\alpha$ & 22.1 & 0.1\\
\\
	&51517.05 & d & B	& 22.2 & 0.1\\
	&51521.04 & f & B	& 23.0 & 0.1\\
	&51523.13 & g & B	& 23.2 & 0.2\\
	&51525.10 & h & B	& $>$23.0 &\nodata\\
\\
4	&51519.07 & e & H$\alpha$ & 22.8 & 0.1\\
	&51523.10 & g & H$\alpha$ & 23.1 & 0.2\\
	&51525.05 & h & H$\alpha$ & 22.8 & 0.1\\
\\
	&51517.05 & d & B	& $>$24.9 &\nodata\\
	&51521.04 & f & B	& 24.8 & 0.3\\
	&51523.13 & g & B	& $>$23.8 &\nodata\\
	&51525.10 & h & B	& $>$23.0 &\nodata\\
\\
5	&52697.07 & i & H$\alpha$ & 22.8 & 0.1\\
\\
6	&51519.07 & e & H$\alpha$ & 24.2 & 0.5\\
	&51523.10 & g & H$\alpha$ & $>$23.5 &\nodata\\
	&51525.05 & h & H$\alpha$ & $>$22.8 &\nodata\\
\\
	&51517.05 & d & B	& 23.4 & 0.1\\
	&51521.04 & f & B	& 24.2 & 0.1\\
	&51523.13 & g & B	& $>$ 24.3 &\nodata\\
	&51525.10 & h & B	& $>$ 22.7 &\nodata
\enddata
\tablenotetext{a}{Photometry for this object was forced to the R-band
system.}
\end{deluxetable}


\end{document}